\documentclass[fleqn,usenatbib]{mnras}
\usepackage{newtxtext,newtxmath}
\usepackage[T1]{fontenc}
\usepackage{ae,aecompl}

\usepackage{graphicx}	% Including figure files
\usepackage{amsmath}	% Advanced maths commands
\usepackage{amssymb}	% Extra maths symbols

\newcommand{\Gaia}{\textit{Gaia}}
\newcommand{\HST}{HST}
\newcommand{\thetaE}{\Theta_{\text{E}}}
\newcommand{\ml}{M_{L}}
\newcommand{\ks}{K$_{\text{s}}$}
\defcitealias{Br18}{B18}
\defcitealias{S17}{S17}
\defcitealias{Z18}{Z18}
\defcitealias{Kl18b}{K18}

\title[VVV and \textit{Gaia} microlensing]{Ongoing astrometric microlensing events from VVV and \textit{Gaia}}

\author[P. McGill et al.]{
P. McGill,$^{1}$\thanks{E-mail: pm625@cam.ac.uk (PM)}
L. C. Smith,$^{1,2}$
N. W. Evans,$^{1}$
V. Belokurov$^{1}$
and P. W. Lucas$^{2}$
\\
$^{1}$Institute of Astronomy, University of Cambridge, Madingley Rd, Cambridge CB3 0HA, UK\\
$^{2}$Centre for Astrophysics Research, University of Hertfordshire, College Lane, Hatfield AL10 9AB, UK\\
}

\date{Accepted 2019 May 19}

\pubyear{2019}

\begin{document}
\label{firstpage}
\pagerange{\pageref{firstpage}--\pageref{lastpage}}
\maketitle

\begin{abstract}
We extend predictive microlensing event searches using the Vista Variables in the Via Lactea survey and the second Gaia data release. We identify two events with maxima in 2019 that require urgent follow-up. First, we predict that the nearby M2 dwarf L 338-152 will align with a background source with a closest approach of $35^{+35}_{-23}$ mas on 2019 November $16^{+28}_{-27}$ d. This will cause a peak astrometric shift and photometric amplification of the background source of $2.7^{+3.5}_{-1.5}$ mas and $5.6^{+143.2}_{-5.2}$ mmag respectively. This event should be astrometrically detectable by both the Hubble Space Telescope (HST) and the Spectro-Polarimetric High-contrast Exoplanet Research instrument on the Very Large Telescope. Secondly, we predict the likely K dwarf NLTT 45128 will lens a background source with a closest approach of $105.3^{+12.2}_{-11.7}$ mas on 2019 September $26^{+15}_{-15}$ d. This will produce a peak astrometric shift of $0.329^{+0.065}_{-0.059}$ mas. NLTT 45128 is only 3.6 magnitudes brighter than the background source which makes it an excellent candidate for follow-up with HST. Characterisation of these signals will allow  direct gravitational masses to be inferred for both L 338-152 and NLTT 45128 with an estimated precision of $\sim9$ and $\sim13$ per cent respectively.
\end{abstract}


%%%%%%%%%%%%%%%%%%%%%%%%%%%%%%%%%%%%%%%%%%%%%%%%%%

%%%%%%%%%%%%%%%%% BODY OF PAPER %%%%%%%%%%%%%%%%%%

\section{Introduction}

Astrometric microlensing events provide unique opportunities to obtain direct gravitational mass determinations of isolated stars and stellar remnants. Recently, this effect was measured for the first time outside our solar system \citep[][hereafter \citetalias{S17}]{S17} using the Hubble Space Telescope (\HST{}). \citetalias{S17} determined the mass of white dwarf Stein 2051B to a precision of $\sim7.5\%$, allowing a direct test of white dwarf mass-radius relationships. This was followed by a mass determination of our closest neighbour Proxima Centauri via astrometric microlensing to $\sim40\%$ precision \citep[][hereafter \citetalias{Z18}]{Z18}, using ground based adaptive optics with the Spectro-Polarimetric High-contrast Exoplanet REsearch \citep[SPHERE;][]{Be19} instrument on the Very Large Telescope (VLT).

For these events to be monitored, they must first be predicted. That is, we must determine when a foreground lens will align closely enough with a background source to produce a detectable microlensing effect. \Gaia{} data has proven ideal for predicting astrometric microlensing events~\citep[][hereafter \citetalias{Br18}]{Mc18,Br18}. In particular, data release 2, which provides excellent astrometric solutions for $\sim 1.7$ billion objects \citep[hereafter GDR2]{GDR2}, has opened up the flood gates. These astrometric solutions have enabled the prediction of future stellar alignments at the required precision for high confidence microlensing event predictions. Searches using solely GDR2 have forecast hundreds of upcoming events both astrometric (\citetalias{Br18}; \citealp{Of18}; \citealp{Kl18a}; \citealp{Kl18b}, hereafter \citetalias{Kl18b}) and for the first time photometric (\citetalias{Br18}; \citealp{Mu18}; \citetalias{Kl18b}; \citealp{Mc19}). Moreover, \cite{Br&N18} demonstrated that the GDR2 astrometric solution is good enough to predict microlensing events over the next century and presented an almanac of 2,509 events.

The only work to date to extend searches beyond GDR2 is \cite{Ne18}. They used a population of potential lenses consisting of nearby very low mass objects from PanSTARRS data release 1, and a background source population from GDR2. The rationale for doing this is to remedy GDR2's incompleteness for faint high proper motion objects. Using this complementary dataset to GDR2, \cite{Ne18} found a further 27 predicted microlensing events over the next 50 years. In this study, we extend predictive microlensing by using a complementary background source population from the Vista Variables in the Via Lactea survey (VVV, \citealt{DM10}) with GDR2 lenses. This allows us to probe deeper (\ks{}-band $\sim$ 17 mag) into areas of high source density in the galactic bulge and southern disk where microlensing events are most likely to happen \citep[e.g.][]{Na18}.

This \textit{Letter} is structured as follows. First, we summarize the observed signals of a microlensing event in which the lens and source can be resolved. Next we outline the method used to find microlensing events using GDR2 and VVV. Finally, we present two candidate events that require immediate attention and assess the outlook for follow-up observations.

\section{Partially resolved microlensing }

When a foreground object (the lens) with mass $\ml$ intervenes between an observer and distant background source (the source), light from the source is gravitationally deflected. In the case of perfect alignment between the lens and source, an Einstein ring of angular radius
\begin{equation}
    \thetaE{} = \sqrt{\frac{4G\ml}{c^{2}}\left(\frac{1}{D_{L}}-\frac{1}{D_{S}}\right)} = 2.854 \sqrt{\frac{\ml}{M_{\sun}}\frac{\varpi_{L}-\varpi_{S}}{\text{mas}}}\quad\text{mas}
    \label{eq:thetaE}
\end{equation}
is formed. Here $D_{L}$,$D_{S}$,$\varpi_{L}$ and $\varpi_{S}$ are the lens and source distances and parallaxes, $c$ is the speed of light, and $G$ is the gravitational constant. During a microlensing event, when a lens aligns imperfectly with a source, two images are formed. The brighter major image (+) is always formed on the outside of the Einstein ring at larger angular separation from the lens, whereas the fainter minor image is formed on the inside of the Einstein ring with smaller angular separation from the lens. 

If the major source image is sufficiently distant from the lens such that it can be resolved, two effects can be observed. First, a transient brightening of the major image is seen with the amplification of the source's unlensed flux given by \cite{P86} as, 
\begin{equation}
    A_{+} = \frac{u^{2}+2}{2u\sqrt{u^2+4}} + \frac{1}{2}.
    \label{eq:phot}
\end{equation}
Here, $u =|\Delta\phi|/\thetaE{}$ is the instantaneous normalised angular separation of the lens and source. This amplification is only detectable for small lens source angular separations ($|\Delta\phi|\sim\thetaE{}$ or $u\sim1$). $A_{+}$ is maximal at closest approach when $|\Delta\phi| =\Delta\phi_{\text{min}}$ or $u = u_{\text{min}}$. 

At much larger lens source separations ($|\Delta\phi| >>\thetaE{}$), an astrometric shift of the source from its unlensed position can also be detected. This shift is given by the position of the major image relative to the unlensed source position during the event as (e.g. \citetalias{S17})
\begin{equation}
    \delta\theta_{+} = \frac{1}{2}\left(\sqrt{u^{2}+4} - u\right)\frac{\thetaE{}}{\text{mas}}\quad\text{mas}.
    \label{eq:ast}
\end{equation}
If $\delta\theta_{+}$ can be measured and the lens and source distances are known, then $M_{L}$ can be directly inferred via eqn~(\ref{eq:thetaE}).

However, if the lens and major source image cannot be resolved during the event, for large normalised separations ($u>>1$), both the photometric and astrometric signals are suppressed by a factor of $\sim1/(1+g)$, where $g$ is the ratio of the lens and source flux \citep{D00}.  For typical predictable microlensing events, nearby high-proper motion objects are considered as potential lenses and therefore the lens is often several magnitudes brighter than the source. This results in a suppression of the signals by factors of $\sim100$, rendering them undetectable. \citetalias{Br18} (especially section 3) provides a comprehensive discussion of the behavior of these signals in the partially resolved and unresolved regimes.

%For a typical predicted event, the lens is often several magnitudes brighter than the source and this results in suppression by factors of $\sim100$, rendering them undetectable.

%For a typical precdicable microlensing event where nearby high-proper motion objects are considered as potential lenses the lens is often several magnitudes brighter than the source. 

\begin{figure}
    \centering
    \includegraphics[width=0.49\linewidth]{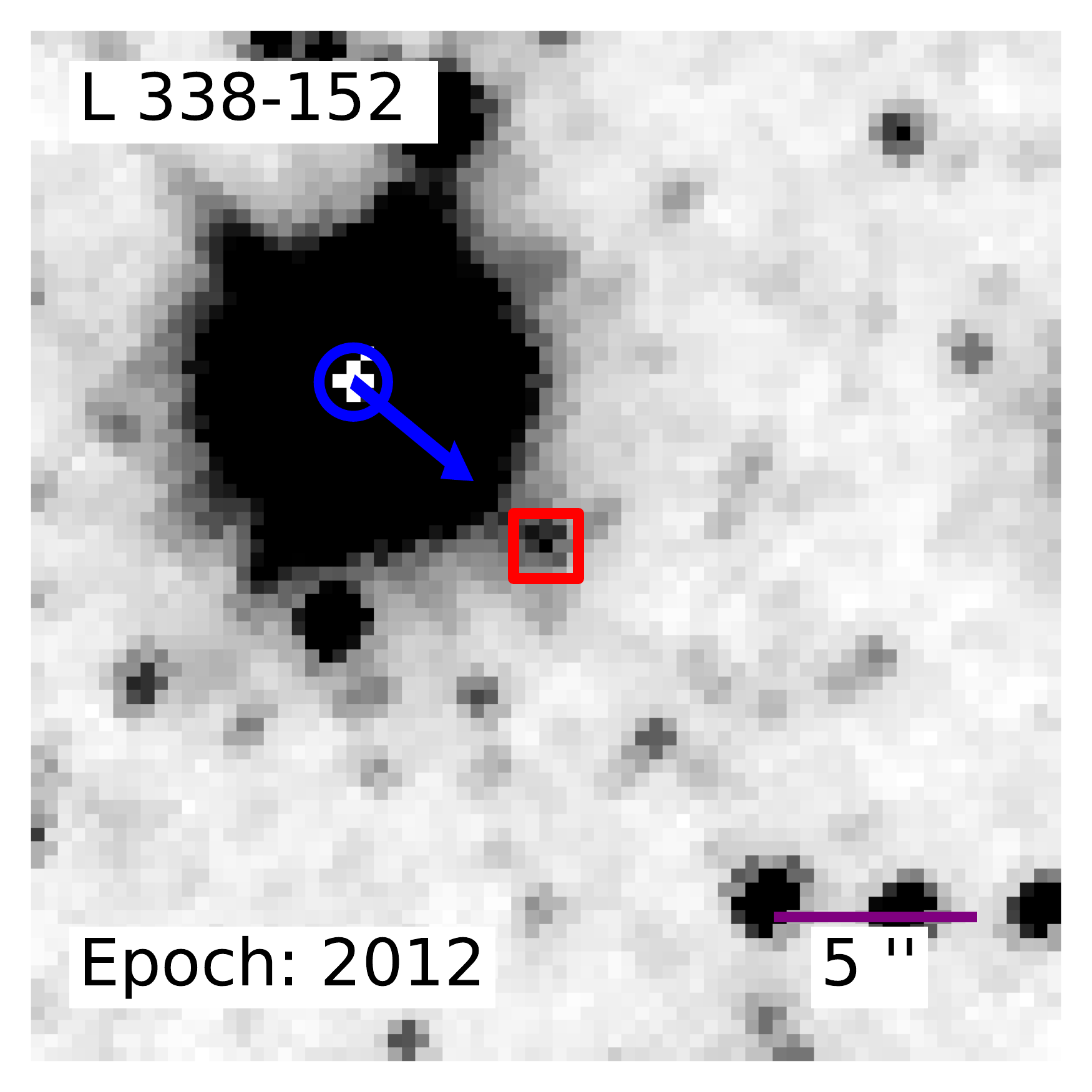}
    \includegraphics[width=0.49\linewidth]{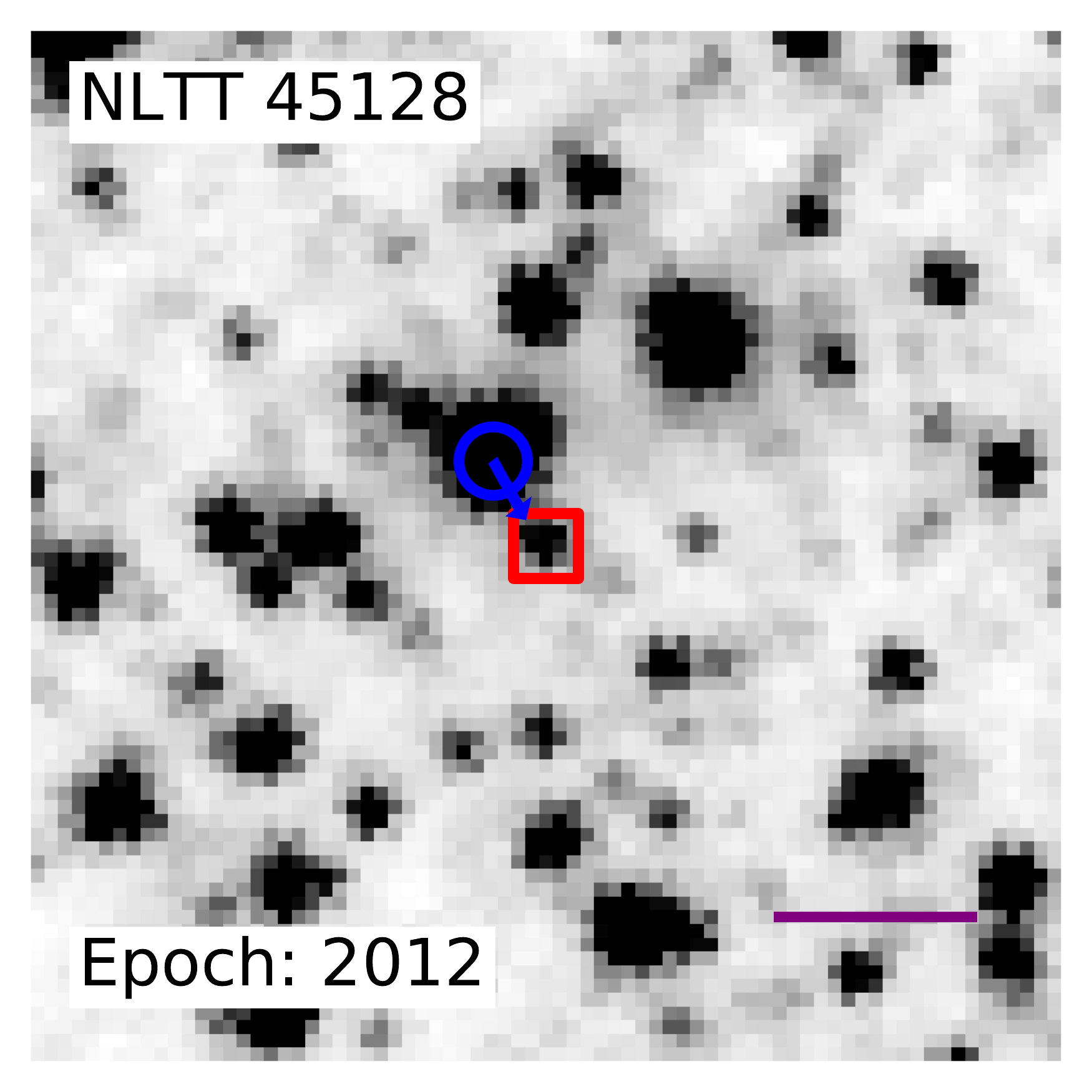}
    \caption{Stellar field around each of the candidate events. In each image, the blue circle and red square indicate the position of the lens and source at the image epoch respectively. The blue arrow is the proper motion vector of the lens and length of 5 years of proper motion. The epoch of each image is indicated in the bottom left in Julian Years, and the SIMBAD lens name is in the top right. Both images are \ks-band cutouts from VVV.}
    \label{fig:cutouts}
\end{figure}

%LCS note that the VVV 'K' band is the short K, or K_s band. I fixed this in the above caption but I suggest you check elsewhere. I added the new command \ks 

\section{Predicting Microlensing Events}

To predict future microlensing events, the position, proper motion, and parallax of both the lens $(\alpha_{0L},\delta_{0L},\mu_{\alpha* L},\mu_{\delta L},\varpi_{L})$ and source $(\alpha_{0S},\delta_{0S},\mu_{\alpha*S},\mu_{\delta S},\varpi_{S})$ are required. This information allows the future on-sky separation between a lens and source to be calculated as seen from Earth as
$\Delta\bphi(t) = \bphi_{L}(t) - \bphi_{S}(t),$
where both $\bphi_{L}(t)$ and $\phi_{S}(t)$ can be found using their respective astrometric parameters as
$\bphi(t) \approx \bphi_{0} +\bmu [t-t_{0}] +\varpi\textbf{P}(t)$
with, 
\begin{equation}
\bphi_{0} =
\begin{pmatrix}
    \alpha_{0} \\
    \delta_{0}
\end{pmatrix},  \qquad\qquad  
\bmu = \begin{pmatrix}
    \mu_{\alpha*}/\cos\delta_{0} \\
    \mu_{\delta}
\end{pmatrix},
\end{equation}
and

\begin{equation}
\mathbfit{P}(t) = 
\begin{pmatrix}
\left[X(t)\sin\alpha_{0}-Y(t)\cos\alpha_{0}\right]/\cos\delta_{0}\\
X(t)\cos\alpha_{0}\sin\delta_{0}+Y(t)\sin\alpha_{0}\sin\delta_{0}-Z(t)\cos\delta_{0}
\end{pmatrix}.
\end{equation}

\noindent
Here $t_{0}$ is the reference epoch and $[X(t),Y(t),Z(t)]$ are the Cartesian barycentric solar system coordinates in AU of the earth on the ICRF at time $t$. They were retrieved via the {\tt astropy} Python package \citep{astropy18}, which uses values computed from NASA JPL's Horizons Ephemeris.
Using an estimate for the lens mass $\ml$, $\thetaE$ can be calculated using eqn~(\ref{eq:thetaE}). This then allows future values of $u$ for a lens and source to be found and hence both the photometric (eqn~\ref{eq:phot}) and astrometric (eqn~\ref{eq:ast}) signals can be predicted.

\section{Search for Predicted Events}

%In this section, we describe how the lens and background source samples were selected from GDR2 and VIRAC respectively. We also describe the procedure for estimating the lens masses and the search method used to find events.

\subsection{The VVV Infrared Astrometric Catalogue (VIRAC)}

The construction of our background source catalogue (VIRAC) is described in detail in Smith et al. (in prep), and deviates significantly from \citet{LCS18}. A brief summary of the most relevant details of the process is provided here. They obtained VVV and VVVx \ks-band images covering the original VVV area that were observed between the start of 2010 and October 2018. These were then processed with \textsc{dophot} \citep{dophot93} to produce source catalogues as described by \citet{AG18}. They apply the astrometric solutions contained in the image headers and then crossmatch each source catalogue to projected GDR2 source positions at the image epoch. At each point on a 10$\times$10 grid placed over each VIRCAM (VISTA InfraRed CAMera) array -- so grid points are therefore approx 1.3\arcmin{} apart -- they fit plate constants using local \Gaia{} reference sources. Linear interpolation of the plate constants between grid points in effect produces a piecewise linear function transforming array coordinates onto the GDR2 coordinate system. They apply the transformation, match sources between overlapping observations, and use a least square fit implemented by the {\tt SciPy} Python package \citep{scipy} to fit either a 4 parameter proper motion and position (at epoch J2014.0) solution when a source has 5 to 19 epochs inclusive, or a 5 parameter solution which also includes parallax when the source has $\geq20$ epochs. If a source has 2 to 4
epochs inclusive, then a simple average position (at the average epoch) is recorded. Crucially for this work, all positions, proper motions and parallaxes are anchored to the GDR2 astrometric reference frame. This catalogue maintains a high completeness approximately 1 to 1.5 \ks{} magnitudes deeper than its predecessor.

\subsection{Selection and lens mass estimates}

To select stars which have a high probability of lensing a background source, we take a high proper motion sample (total proper motion > 150 mas yr$^{-1}$) of stars from GDR2 that lie within the VVV footprint. To remove spurious high proper motion objects, we insist on high quality astrometric solutions. Specifically, we employ a magnitude dependent cut on the unit weight error of an object's astrometric solution defined and suggested by Appendix C, equation C1 of \cite{L18}. We also require that a lens has a positive parallax that is less than that of the closest star Proxima Centauri ($0<\varpi_{L}<768.5$ mas). Finally, we require high significance parallax detection
$\varpi_{L}/\sigma_{\varpi_{L}} > 10$, which ensures that a high precision lens mass can be extracted from the astrometric signal. This leaves a sample of 1915 lens stars ranging from $\sim$ 2-21 in G-band apparent magnitude, of which 1638 have measured $G_{\text{BP}}-G_{\text{RP}}$ colors. For the selection of the background source sample from VIRAC we require a strict quality cut. We require a source to be detected for $> 6$ consecutive calendar years in the best seeing pointing covering the source each year. This leaves a sample of 436\,903\,339 source stars ($\sim$ 25$\%$ of the total number of VIRAC sources).

%This is because spurious objects are likely to be found near bright objects. As we are selecting a source close to a likely bright high proper motion GDR2 lens, this unfortunately makes us prone to picking these spurious objects. We therefore impose strict quality cuts on VIRAC sources. We require a source to be detected for $> 6$ consecutive calendar years in the best seeing pointing covering the source each year. This leaves a sample of 436\,903\,339 source stars.

We estimate the mass of each lens using the method suggested by \citetalias{Kl18b}. Specifically, we first classify each lens as either main sequence, red giant, white dwarf, or brown dwarf according to \Gaia{} $G$, $G_{\text{BP}}$, $G_{\text{RP}}$ magnitudes and parallax. If a lens has no $G_{\text{BP}}$, or $G_{\text{RP}}$ measurements, it is assumed to be on the main sequence. For red giants, white dwarfs and brown dwarfs, we assume fixed masses of ($1.0\pm0.5,0.65\pm0.15,0.07\pm0.03$) $M_{\sun}$ respectively. For main sequence lenses, we use \citetalias{Kl18b}'s empirical mass luminosity relations and assume a 10 per cent error. After an event is found, we update the mass, if a more accurate measurement derived using spectroscopy and stellar models is available from the literature.

%LCS I wonder if K18 or others were asked (or you will be asked) by the referee whether the hit-and-miss Bp and Rp measurements in these crowded regions are influencing your selection. Perhaps worth thinking about even if not commented on at this stage.

\begin{figure*}
    \centering
    \includegraphics[width=\linewidth]{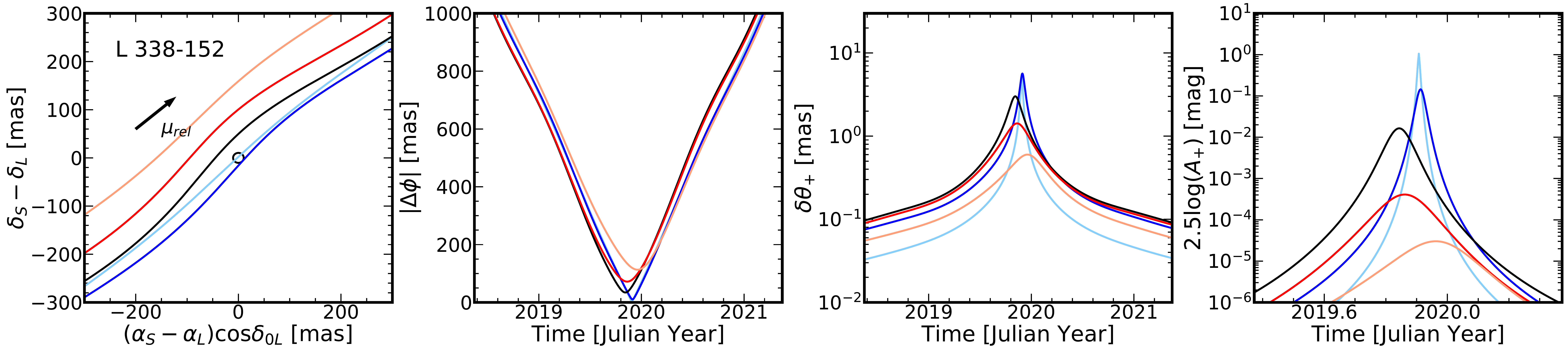}
    \includegraphics[width=\linewidth]{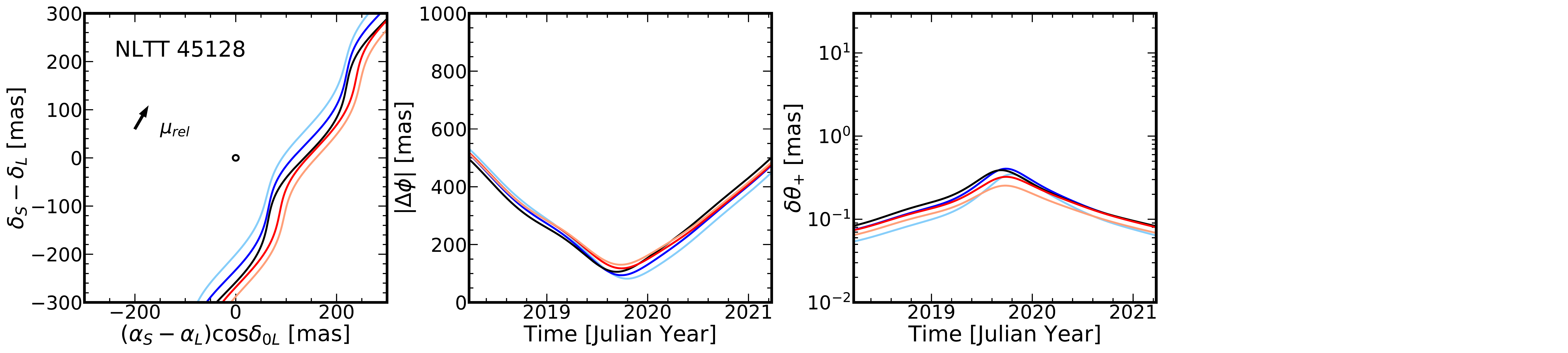}
    \caption{Predicted trajectories and signals for the microlensing event candidates. In each panel the light-blue, blue, black, red and light-red curves are example trajectories and signals corresponding to the the  2.3, 15.9, 50, 84.1, and 97.7 percentiles, respectively, of $u_\text{min}$ calculated in the Monte Carlo simulations described in section 4.3. (Left to right) The first column shows samples of the predicted source trajectory in the lens rest frame. The median of the estimated angular Einstein ring is shown as a black circle. SIMBAD names for the lenses are also indicated for each event. The relative proper motion vector $\bmu_{\textbf{rel}}=\bmu_{S}-\bmu_{L}$ is shown with a length of 0.l years of motion. The second column shows samples of the predicted separation of the lens and source around the event maximum. The third column shows samples of the predicted resolved astrometric shift of the source. Finally, the fourth column shows samples of the predicted photometric amplification for the only event with a detectable peak signal > 1 mmag.}
    \label{fig:signals}
\end{figure*}

\subsection{Search method}

To search for predicted events among the large number of possible pairings of GDR2 lenses and VIRAC sources, we only compute predicted signals for lenses that pass sufficiently close to background sources for detectable signals to be produced. To achieve this we use a method similar to \cite{P11} and \citetalias{Kl18b}. Specifically, we match each lens with all sources that pass within a distance of 10\arcsec{} of the predicted lens trajectory over the next $\sim$20 years (2018-2040). This leaves 7827 lens source pairs to be checked. The 10\arcsec{} match distance chosen is more than sufficient to match all sources that are deflected by a detectable amount ($> 0.2$ mas) by a typical nearby stellar lens with $\thetaE \sim 10$ mas.

Next, for each lens source pair, we find the time of minimum separation by numerically minimizing the angular separation between them. In some cases, parallax motion causes the angular separation to have local minima. In order to find the global minimum, we first minimize the angular separation without parallax motion (which can be done analytically), and then use this as an initialization to the basin-hopping algorithm \citep{w97} implemented by {\tt SciPy}. The basin-hopping algorithm combines global stepping with local minizations allowing the global minimum to be found. We set the temperature parameter for the basin-hopping algorithm (expected distance between local minima) to 0.5 years.

Using the estimated $\thetaE{}$, we calculate the peak astrometric and photometric signals. Uncertainties are derived using Monte Carlo simulations. We draw $10^{4}$ samples from Gaussian distributions consistent with the means and covariances reported by GDR2 and VIRAC for the astrometric parameters, and our lens mass estimates. VIRAC parallax measurement uncertainties are often at the level of a few mas, particularly in the vicinity of bright sources as is the case for all our candidates. Because sources in the VVV area are generally several kpc distant, many scatter below zero in measured parallax. Hence, we draw from Gaussian distributions truncated at 0, which avoids unphysical negative source parallax values.

For an event to be astrometrically detectable, we require a peak astrometric shift $> 0.2$ mas. For a photometric event, we require a peak amplification of $> 1.0$ mmag. This is in line with the current instrumentation detection capabilities (e.g. with \HST{}, \Gaia{} and SPHERE: \citetalias{S17}; \citealp{Ka18}; \citealp{Kr18}; \citetalias{Z18}). 
This leaves 14 candidate events with maxima in 2019. Visual inspection of Digitized Sky Surveys 2 and VVV \ks-band image data eliminated 12 events. In these cases, only the VIRAC source and not the lens could be definitively identified in the image. To avoid duplicating the efforts of other searches using GDR2 as background sources, we remove events where a VIRAC source lies within 1.0\arcsec{} from a GDR2 detection at epoch J2014.0. This cut has no effect, leaving two new candidate events.

\section{Candidate Events}

\begin{table}
\caption{Details of the microlensing events. Lens and source astrometric parameters are on the IRCF and obtained from GDR2 and VIRAC respectively. Lens and source initial positions are at epochs J2015.5 and J2014.0 respectively. \ks-band magnitude of the lens and source are from 2MASS and VIRAC respectively. Subscripts L and S represent parameters of the lens and source respectively. $T[\delta\theta_{+},A_{+}]$ are the full-width-half-maximums of the signals. () indicate uncertainty on the last digit. Median $\pm$ 34 percentiles are shown for the event parameters.}
\begin{tabular}{lllll}
\hline
Lens & L 338-152 & NLTT 45128\\
\hline
Type &  M2  & K \\
$\ml{}(M_{\sun{}})$ & 0.51(6) & 0.29(3) \\
$\Theta_{\text{E}}$ (mas)& $10.4^{+1.8}_{-2.7}$ & $5.94^{+0.20}_{-0.35}$ \\
$\delta\theta_{+}$(mas) & $2.7^{+3.5}_{-1.5}$ & $0.329^{+0.065}_{-0.059}$  \\
$A_{+}$(mmag) & $5.6^{+143.2}_{-5.2}$ & - \\ 
$t_{\text{min}}$(Jyr) & $2019.872^{+0.076}_{-0.073}$ & $2019.732^{+0.040}_{-0.040}$ \\
$\Delta\bphi_{\text{min}}$(mas) & $35^{+35}_{-23}$ & $105.3^{+12.2}_{-11.7}$ \\
$u_{\text{min}}$ & $3.5^{+4.0}_{-2.4}$ & $17.9^{+2.7}_{-2.3}$\\
$T[\delta\theta_{+}]$ (d) &  $65^{+57}_{-36}$& $342.9^{+36}_{-32}$ \\
$T[A_{+}]$ (d) & $25^{+24}_{-15}$& - \\
$\varpi_{L}$(mas) & 42.29(6) & 	16.47(5)\\
$\mu_{\alpha*L}$ & -586.9(1) & -158.5(1)\\
(mas/yr) & &  \\
$\mu_{\delta L}$(mas/yr) & -484.55(5) & -286.42(7) \\
$\alpha_{0L}$(deg) & 245.66683036(2) & 264.32271595(1) \\
$\delta_{0L}$(deg) & -48.657646525(9) & -31.22567513(1)\\
\ks$_{L}$(mag) & 7.64(2) & 10.83(3)\\
id$_{L}$(GDR2) & 5941478335062267392 & 4055355782158028416 \\ 
$\varpi_{S}$(mas) & 14(13) & -1(3) \\
$\mu_{\alpha*S}$ & 18(6) & -4(1) \\
(mas/yr) & & \\
$\mu_{\delta S}$(mas/yr) & 30(8) & -8(2) \\
$\alpha_{0S}$(deg) & 245.665710(3) & 264.3224710(7)  \\
$\delta_{0S}$(deg) & -48.658284(5) & -31.2259846(9)\\
\ks$_{S}$ (mag) & 15.638(9) & 14.452(2) \\

%id$_{S}$(VIRAC) & 688911062303 &  597232099277 \\
%LCS commented out the virac2 ids, they will be meaningless to anyone else for several months and may change (although that's unlikely). The position should be more than adequate to identify them in virac2. We can leave them commented in the tex source so any intrepid explorers can find them.

\hline
\end{tabular}
\label{tab:eventdetail}
\end{table}

\subsection{L 338-152}

L 338-152 is a 7.6th magnitude (\ks-band) M2 dwarf \citep{Ra13} with a GDR2 parallax and total proper motion of $42.2$ mas, $761$ mas/yr. We adopt the mass of $(0.51\pm0.06)$ $M_{\sun}$ for L 338-152 estimated by \cite{Ga14}. This mass was derived by fitting stellar models to L 338-152's spectrum and using empirical mass relations for M dwarfs. We note the mass obtained using \citetalias{Kl18b}'s method of ($0.43\pm0.04$) $M_{\sun}$ provides a reasonable estimate.
We predict L 338-152 will lens a 15.6th magnitude (\ks-band) background source (Fig. \ref{fig:cutouts}). The closest approach of $35^{+35}_{-23}$ mas will occur on J$2019.872^{+0.076}_{-0.073}$ (or 2019 November $16^{+28}_{-27}$ d), with peak resolved shift and magnification of $2.7^{+3.5}_{-1.5}$ mas and $5.6^{+143.2}_{-5.2}$ mmag respectively (see Table \ref{tab:eventdetail} and Fig. \ref{fig:signals}).

\subsection{NLTT 45128}

NLTT 45128 is a little-studied 10.8th magnitude (\ks-band) likely K dwarf star (identified from \Gaia{} $G_{BP}-G_{RP}$ colour and G-band absolute magnitude). It has a GDR2 parallax and total proper motion of 16 mas and 327 mas yr$^{-1}$ respectively. We adopt a mass of ($0.29\pm0.03$)$M_{\sun}$ for NLTT 45128 derived from \citetalias{Kl18b}'s mass luminosity relations. With this in hand, we predict NLTT 45128 will lens a 14.5th (\ks-band) source (Fig \ref{fig:cutouts}) with a closest approach of $105.3^{+12.2}_{-11.7}$ mas on J$2019.732^{+0.040}_{-0.040}$ or (2019 September $26^{+15}_{-15}$ d). This will cause a resolved astrometric shift of the source of $0.329^{+0.065}_{-0.059}$ mas (Table \ref{tab:eventdetail} and Fig. \ref{fig:signals}). As the closest approach of the lens and source is much larger than the Einstein radius ($u\sim18>>1$), there will be no detectable photometric signal.

\subsection{Observational Outlook}

For both events, the background source is not present in GDR2. We thus consider other instruments that have achieved sub-mas single epoch astrometric precision and can observe these southern targets, namely the \HST{}  and SPHERE.

\citetalias{S17} demonstrated that, through single-epoch pointed imaging with the Wide Field Camera 3 (WFC3), \HST{} can achieve a precision of $\sigma_{HST}\sim0.13$ mas. This was attained for a background source $\sim6.5$ mags fainter than the white dwarf lens. \citetalias{S17} notes that the background source could only be resolved at separations $> 450$ mas for this lens source contrast ratio. The predicted minimum separation between Stein 2051B and the source was $\sim203$ mas, meaning that \citetalias{S17} was only able to detect the tails of the deflection.

Using the infrared dual imaging and spectrograph plus integral field spectrograph (IRDIFS) observing mode of SPHERE on the VLT, \citetalias{Z18} achieved a single-epoch astrometric precision of $\sigma_{VLT}\sim0.5$ mas. This was achieved for a background source $\sim$ 11 magnitudes fainter than the lens (Proxima Centauri). SHPERE was able to resolve Proxima Centauri and the background source over the duration of the event, for which the closest separation was $\sim 500$ mas.

L 338-152 will lens a background source that is $\sim 8$ magnitudes fainter, at a closest approach of $\sim 30$ mas. This is a larger contrast ratio than the Stein 2051B event in \citetalias{S17}. Assuming the likely best-case with \HST{} where L 338-152 and the background source could be resolved at $\sim 450$ mas, the astrometric shift would be $\sim 0.24$ mas, which is $\sim  1.8\sigma_{HST}$. This means \HST{} is only likely to detect the astrometric tails of this event. Detection of the photometric signal would require resolution of the lens and source at $\sim 30$ mas separation, which for this lens-source contrast ratio is beyond \HST{}'s capabilities. However, L 338-152 is a bright enough target for good performance with SPHERE. The L 338-152 event has a more favourable lens source contrast ratio than the event successfully detected in \citetalias{Z18} (8 compared to 11 mags difference). However, L 338-152 and the background source would need to be resolved at closer angular separations ($\sim 30$ compared to $\sim 500$ mas). Recent work by \cite{C19} on the characterisation of a faint low-mass companion of HD 142527 shows that for a contrast ratio of $\sim 8$ magnitudes, IRDIFS observing modes can resolve separations of $\sim50-100$ mas. Assuming that L 338-152 and the background source can be resolved at 75 mas, the astrometric deflection would be $\sim 1.41$ mas, which is $\sim2.8\sigma_{VLT}$.
L 338-152 is not visible at low airmass (<1.4) from the VLT at Cerro Paranal between September 2019 to the end of January 2020 (J2019.7-J2020.1). Unfortunately, this is at the maximum of the event at $\sim$ 2019 November 15th, and long enough that the short photometric signal will be missed. However, when L 338-152 becomes visible at low airmass after J2021.1, the background source will be separated by $\sim 100$ mas with an astrometric shift of $\sim 1$ mas which is $\sim2\sigma_{VLT}.$ We conclude that only the astrometric signal will be detectable with SPHERE. 

NLTT 45128 will lens a background source only $\sim3.6$ mag fainter. This contrast ratio is considerably better than the event in \citetalias{S17} (6.5 mags). This means \HST{} will likely be able to resolve NLTT 45128 and the background source for separations $<< 450$ mas. If this event is resolved over closest approach ($\sim105$) mas, it will be detectable at the $\sim 2.5\sigma_{HST}$ level. NLTT 45128 is on the faint end of targets for good performance with SPHERE, and moreover the maximum astrometric shift is $0.327 < \sigma_{VLT}$. We therefore conclude that WFC3 on \HST{} is likely the best instrument for follow-up observation of the NLTT 45128 event

To obtain rough estimates of the precision of the inferred mass from these events, we use Monte Carlo simulations. Inverting eqn (\ref{eq:ast}), the mass of the lens may be written in terms of $\delta\theta_{+},\Delta\phi_{\text{min}},\varpi_{L}$ and $\varpi_{S}$. Assuming the deflection is measured at maximum for L  338-152 and NLTT  45128, we take $\delta\theta_{+}$ to be a Gaussian centred on the median value reported in Table~\ref{tab:eventdetail}, with standard deviation of the measurement uncertainties of $\sigma_{VLT}$ and $\sigma_{HST}$ respectively. We then take Gaussian distributions consistent with the GDR2 value for $\varpi_{L}$, and for $\Delta\phi_{\text{min}},\varpi_{S}$, we centre on the medians reported in Table \ref{tab:eventdetail} but with standard deviations of 1 mas, and truncated at 0 for $\varpi_{S}$. Here, we assumed that the source's astrometric solution can be refined by space-based observations (e.g with HST at baseline after the event). Taking $10^5$ samples for both events per epoch, neglecting covariances, and assuming 8 epochs of data (as was achieved in \citetalias{S17}) can be taken around maximum (with $1/\sqrt{8}$ error scaling over all epochs), we recover the mass of L  338-152 and NLTT  45128 to $\sim 9$ and $\sim 13$ per cent respectively.

\section{Conclusions}

We have extended predictive microlensing searches beyond \Gaia{} using the VIRAC catalogue derived from observations from the VVV. This allowed us to probe deep (\ks{}-band $\sim$ 17 mag) into regions of high source density in the galactic bulge and southern disk, and find events missed by previous searches. We identify two events with detectable signals, with maxima in 2019, that require immediate follow-up.

We predict that L 338-152, a M2 star, will lens a background source on 2019 November $16^{+28}_{-27}$ d. This will produce a resolved astrometric shift of $2.7^{+3.5}_{-1.5}$ mas and resolved photometric amplification of $5.6^{+143.2}_{-5.2}$ mmag. Although the photometric signal of this event is undetectable with the currently available instrumentation, the astrometric shift of the background source is likely to be detectable by instruments on both \HST{} and the VLT. We also predict that NLTT 45128, a likely K-dwarf, will lens a background source on 2019 September $26^{+15}_{-15}$ d. This will produce a maximum astrometric shift of the background source of $0.329^{+0.065}_{-0.059}$ mas. This event has a particularly favourable contrast ratio (the source is only 3.6 mags fainter than NLTT 45128) and the shift will likely be detectable with the \HST{}. Characterisation of these astrometric signals will allow direct mass determinations of L 338-152 and NLTT 45128 to $\sim9$ and $\sim13$ per cent precision.

\section*{Acknowledgements}

PM would like to thank STFC for studentship funding and Sergey Koposov for use of the wsdb. We would like to thank the referee Daniel Bramich for a review that greatly improved the paper. PM would also like to thank Felicia Gutmans for carefully reading the paper. This work has made use of data from the European Space Agency (ESA) mission {\it Gaia} \url{https://www.cosmos.esa.int/gaia}, processed by
the {\it Gaia} Data Processing and Analysis Consortium (DPAC \url{https://www.cosmos.esa.int/web/gaia/dpac/consortium}). Funding for the DPAC has been provided by national institutions, in particular the institutions participating in the {\it Gaia} Multilateral Agreement.
%%%%%%%%%%%%%%%%%%%%%%%%%%%%%%%%%%%%%%%%%%%%%%%%%%


\bibliographystyle{mnras}
\bibliography{paper}

\begin{thebibliography}{}
\makeatletter
\relax
\def\mn@urlcharsother{\let\do\@makeother \do\$\do\&\do\#\do\^\do\_\do\%\do\~}
\def\mn@doi{\begingroup\mn@urlcharsother \@ifnextchar [ {\mn@doi@}
  {\mn@doi@[]}}
\def\mn@doi@[#1]#2{\def\@tempa{#1}\ifx\@tempa\@empty \href
  {http://dx.doi.org/#2} {doi:#2}\else \href {http://dx.doi.org/#2} {#1}\fi
  \endgroup}
\def\mn@eprint#1#2{\mn@eprint@#1:#2::\@nil}
\def\mn@eprint@arXiv#1{\href {http://arxiv.org/abs/#1} {{\tt arXiv:#1}}}
\def\mn@eprint@dblp#1{\href {http://dblp.uni-trier.de/rec/bibtex/#1.xml}
  {dblp:#1}}
\def\mn@eprint@#1:#2:#3:#4\@nil{\def\@tempa {#1}\def\@tempb {#2}\def\@tempc
  {#3}\ifx \@tempc \@empty \let \@tempc \@tempb \let \@tempb \@tempa \fi \ifx
  \@tempb \@empty \def\@tempb {arXiv}\fi \@ifundefined
  {mn@eprint@\@tempb}{\@tempb:\@tempc}{\expandafter \expandafter \csname
  mn@eprint@\@tempb\endcsname \expandafter{\@tempc}}}

\bibitem[\protect\citeauthoryear{{Alonso-Garc{\'{\i}}a}
  et~al.,}{{Alonso-Garc{\'{\i}}a} et~al.}{2018}]{AG18}
{Alonso-Garc{\'{\i}}a} J.,  et~al., 2018, \mn@doi [\aap]
  {10.1051/0004-6361/201833432}, \href
  {http://adsabs.harvard.edu/abs/2018A%26A...619A...4A} {619, A4}

\bibitem[\protect\citeauthoryear{{Astropy Collaboration} et~al.,}{{Astropy
  Collaboration} et~al.}{2018}]{astropy18}
{Astropy Collaboration} et~al., 2018, \mn@doi [\aj] {10.3847/1538-3881/aabc4f},
  \href {http://adsabs.harvard.edu/abs/2018AJ....156..123A} {156, 123}

\bibitem[\protect\citeauthoryear{{Beuzit} et~al.,}{{Beuzit}
  et~al.}{2019}]{Be19}
{Beuzit} J.~L.,  et~al., 2019, arXiv e-prints, \href
  {https://ui.adsabs.harvard.edu/\#abs/2019arXiv190204080B} {p.
  arXiv:1902.04080}

\bibitem[\protect\citeauthoryear{{Bramich}}{{Bramich}}{2018}]{Br18}
{Bramich} D.~M.,  2018, \mn@doi [\aap] {10.1051/0004-6361/201833505}, \href
  {https://ui.adsabs.harvard.edu/\#abs/2018A&A...618A..44B} {618, A44}

\bibitem[\protect\citeauthoryear{{Bramich} \& {Nielsen}}{{Bramich} \&
  {Nielsen}}{2018}]{Br&N18}
{Bramich} D.~M.,  {Nielsen} M.~B.,  2018, \mn@doi [\actaa]
  {10.32023/0001-5237/68.3.1}, \href
  {https://ui.adsabs.harvard.edu/\#abs/2018AcA....68..183B} {68, 183}

\bibitem[\protect\citeauthoryear{{Claudi} et~al.,}{{Claudi} et~al.}{2019}]{C19}
{Claudi} R.,  et~al., 2019, \mn@doi [\aap] {10.1051/0004-6361/201833990}, \href
  {https://ui.adsabs.harvard.edu/\#abs/2019A&A...622A..96C} {622, A96}

\bibitem[\protect\citeauthoryear{{Dominik} \& {Sahu}}{{Dominik} \&
  {Sahu}}{2000}]{D00}
{Dominik} M.,  {Sahu} K.~C.,  2000, \mn@doi [\apj] {10.1086/308716}, \href
  {https://ui.adsabs.harvard.edu/\#abs/2000ApJ...534..213D} {534, 213}

\bibitem[\protect\citeauthoryear{{Gaia Collaboration} et~al.,}{{Gaia
  Collaboration} et~al.}{2018}]{GDR2}
{Gaia Collaboration} et~al., 2018, \mn@doi [\aap]
  {10.1051/0004-6361/201833051}, \href
  {https://ui.adsabs.harvard.edu/\#abs/2018A&A...616A...1G} {616, A1}

\bibitem[\protect\citeauthoryear{{Gaidos} et~al.,}{{Gaidos}
  et~al.}{2014}]{Ga14}
{Gaidos} E.,  et~al., 2014, \mn@doi [\mnras] {10.1093/mnras/stu1313}, \href
  {https://ui.adsabs.harvard.edu/\#abs/2014MNRAS.443.2561G} {443, 2561}

\bibitem[\protect\citeauthoryear{Jones, Oliphant, Peterson  et~al.}{Jones
  et~al.}{2001}]{scipy}
Jones E.,  Oliphant T.,  Peterson P.,   et~al., 2001, {SciPy}: Open source
  scientific tools for {Python}, \url {http://www.scipy.org/}

\bibitem[\protect\citeauthoryear{{Kains}, {Calamida}, {Sahu}, {Anderson},
  {Casertano}  \& {Bramich}}{{Kains} et~al.}{2018}]{Ka18}
{Kains} N.,  {Calamida} A.,  {Sahu} K.~C.,  {Anderson} J.,  {Casertano} S.,
  {Bramich} D.~M.,  2018, \mn@doi [\apj] {10.3847/1538-4357/aae311}, \href
  {http://adsabs.harvard.edu/abs/2018ApJ...867...37K} {867, 37}

\bibitem[\protect\citeauthoryear{{Kl{\"u}ter}, {Bastian}, {Demleitner}  \&
  {Wambsganss}}{{Kl{\"u}ter} et~al.}{2018a}]{Kl18a}
{Kl{\"u}ter} J.,  {Bastian} U.,  {Demleitner} M.,   {Wambsganss} J.,  2018a,
  \mn@doi [\aap] {10.1051/0004-6361/201833461}, \href
  {https://ui.adsabs.harvard.edu/\#abs/2018A&A...615L..11K} {615, L11}

\bibitem[\protect\citeauthoryear{{Kl{\"u}ter}, {Bastian}, {Demleitner}  \&
  {Wambsganss}}{{Kl{\"u}ter} et~al.}{2018b}]{Kl18b}
{Kl{\"u}ter} J.,  {Bastian} U.,  {Demleitner} M.,   {Wambsganss} J.,  2018b,
  \mn@doi [\aap] {10.1051/0004-6361/201833978}, \href
  {https://ui.adsabs.harvard.edu/\#abs/2018A&A...620A.175K} {620, A175}

\bibitem[\protect\citeauthoryear{{Lindegren} et~al.,}{{Lindegren}
  et~al.}{2018}]{L18}
{Lindegren} L.,  et~al., 2018, \mn@doi [\aap] {10.1051/0004-6361/201832727},
  \href {https://ui.adsabs.harvard.edu/\#abs/2018A&A...616A...2L} {616, A2}

\bibitem[\protect\citeauthoryear{{McGill}, {Smith}, {Evans}, {Belokurov}  \&
  {Smart}}{{McGill} et~al.}{2018}]{Mc18}
{McGill} P.,  {Smith} L.~C.,  {Evans} N.~W.,  {Belokurov} V.,   {Smart} R.~L.,
  2018, \mn@doi [\mnras] {10.1093/mnrasl/sly066}, \href
  {http://adsabs.harvard.edu/abs/2018MNRAS.478L..29M} {478, L29}

\bibitem[\protect\citeauthoryear{{McGill}, {Smith}, {Evans}, {Belokurov}  \&
  {Zhang}}{{McGill} et~al.}{2019}]{Mc19}
{McGill} P.,  {Smith} L.~C.,  {Evans} N.~W.,  {Belokurov} V.,   {Zhang} Z.~H.,
  2019, \mn@doi [\mnras] {10.1093/mnras/sty3344}, \href
  {https://ui.adsabs.harvard.edu/\#abs/2019MNRAS.483.4210M} {483, 4210}

\bibitem[\protect\citeauthoryear{{Minniti} et~al.,}{{Minniti}
  et~al.}{2010}]{DM10}
{Minniti} D.,  et~al., 2010, \mn@doi [\na] {10.1016/j.newast.2009.12.002},
  \href {http://adsabs.harvard.edu/abs/2010NewA...15..433M} {15, 433}

\bibitem[\protect\citeauthoryear{{Mustill}, {Davies}  \& {Lindegren}}{{Mustill}
  et~al.}{2018}]{Mu18}
{Mustill} A.~J.,  {Davies} M.~B.,   {Lindegren} L.,  2018, \mn@doi [\aap]
  {10.1051/0004-6361/201833527}, \href
  {https://ui.adsabs.harvard.edu/\#abs/2018A&A...617A.135M} {617, A135}

\bibitem[\protect\citeauthoryear{{Navarro}, {Minniti}  \&
  {Contreras-Ramos}}{{Navarro} et~al.}{2018}]{Na18}
{Navarro} M.~G.,  {Minniti} D.,   {Contreras-Ramos} R.,  2018, \mn@doi [\apjl]
  {10.3847/2041-8213/aae08a}, \href
  {http://adsabs.harvard.edu/abs/2018ApJ...865L...5N} {865, L5}

\bibitem[\protect\citeauthoryear{{Nielsen} \& {Bramich}}{{Nielsen} \&
  {Bramich}}{2018}]{Ne18}
{Nielsen} M.~B.,  {Bramich} D.~M.,  2018, \mn@doi [\actaa]
  {10.32023/0001-5237/68.4.3}, \href
  {https://ui.adsabs.harvard.edu/\#abs/2018AcA....68..351N} {68, 351}

\bibitem[\protect\citeauthoryear{{Ofek}}{{Ofek}}{2018}]{Of18}
{Ofek} E.~O.,  2018, \mn@doi [\apj] {10.3847/1538-4357/aadfeb}, \href
  {https://ui.adsabs.harvard.edu/\#abs/2018ApJ...866..144O} {866, 144}

\bibitem[\protect\citeauthoryear{{Paczynski}}{{Paczynski}}{1986}]{P86}
{Paczynski} B.,  1986, \mn@doi [\apj] {10.1086/164140}, \href
  {http://adsabs.harvard.edu/abs/1986ApJ...304....1P} {304, 1}

\bibitem[\protect\citeauthoryear{{Proft}, {Demleitner}  \&
  {Wambsganss}}{{Proft} et~al.}{2011}]{P11}
{Proft} S.,  {Demleitner} M.,   {Wambsganss} J.,  2011, \mn@doi [\aap]
  {10.1051/0004-6361/201117663}, \href
  {https://ui.adsabs.harvard.edu/\#abs/2011A&A...536A..50P} {536, A50}

\bibitem[\protect\citeauthoryear{{Rajpurohit}, {Reyl{\'e}}, {Allard},
  {Homeier}, {Schultheis}, {Bessell}  \& {Robin}}{{Rajpurohit}
  et~al.}{2013}]{Ra13}
{Rajpurohit} A.~S.,  {Reyl{\'e}} C.,  {Allard} F.,  {Homeier} D.,  {Schultheis}
  M.,  {Bessell} M.~S.,   {Robin} A.~C.,  2013, \mn@doi [\aap]
  {10.1051/0004-6361/201321346}, \href
  {https://ui.adsabs.harvard.edu/\#abs/2013A&A...556A..15R} {556, A15}

\bibitem[\protect\citeauthoryear{{Rybicki}, {Wyrzykowski}, {Klencki}, {de
  Bruijne}, {Belczy{\'n}ski}  \& {Chru{\'s}li{\'n}ska}}{{Rybicki}
  et~al.}{2018}]{Kr18}
{Rybicki} K.~A.,  {Wyrzykowski} {\L}.,  {Klencki} J.,  {de Bruijne} J.,
  {Belczy{\'n}ski} K.,   {Chru{\'s}li{\'n}ska} M.,  2018, \mn@doi [\mnras]
  {10.1093/mnras/sty356}, \href
  {https://ui.adsabs.harvard.edu/\#abs/2018MNRAS.476.2013R} {476, 2013}

\bibitem[\protect\citeauthoryear{{Sahu} et~al.,}{{Sahu} et~al.}{2017}]{S17}
{Sahu} K.~C.,  et~al., 2017, \mn@doi [Science] {10.1126/science.aal2879}, \href
  {https://ui.adsabs.harvard.edu/\#abs/2017Sci...356.1046S} {356, 1046}

\bibitem[\protect\citeauthoryear{{Schechter}, {Mateo}  \& {Saha}}{{Schechter}
  et~al.}{1993}]{dophot93}
{Schechter} P.~L.,  {Mateo} M.,   {Saha} A.,  1993, \mn@doi [\pasp]
  {10.1086/133316}, \href {http://adsabs.harvard.edu/abs/1993PASP..105.1342S}
  {105, 1342}

\bibitem[\protect\citeauthoryear{{Smith} et~al.,}{{Smith} et~al.}{2018}]{LCS18}
{Smith} L.~C.,  et~al., 2018, \mn@doi [\mnras] {10.1093/mnras/stx2789}, \href
  {http://adsabs.harvard.edu/abs/2018MNRAS.474.1826S} {474, 1826}

\bibitem[\protect\citeauthoryear{{Wales} \& {Doye}}{{Wales} \&
  {Doye}}{1997}]{w97}
{Wales} D.~J.,  {Doye} J.~P.~K.,  1997, \mn@doi [Journal of Physical Chemistry
  A] {10.1021/jp970984n}, \href
  {http://adsabs.harvard.edu/abs/1997JPCA..101.5111W} {101, 5111}

\bibitem[\protect\citeauthoryear{{Zurlo} et~al.,}{{Zurlo} et~al.}{2018}]{Z18}
{Zurlo} A.,  et~al., 2018, \mn@doi [\mnras] {10.1093/mnras/sty1805}, \href
  {http://adsabs.harvard.edu/abs/2018MNRAS.480..236Z} {480, 236}

\makeatother
\end{thebibliography}

%%%%%%%%%%%%%%%%%%%%%%%%%%%%%%%%%%%%%%%%%%%%%%%%%%

%%%%%%%%%%%%%%%%% APPENDICES %%%%%%%%%%%%%%%%%%%%%

%%%%%%%%%%%%%%%%%%%%%%%%%%%%%%%%%%%%%%%%%%%%%%%%%%

% Don't change these lines
\bsp	% typesetting comment
\label{lastpage}
\end{document}